\documentclass{appolb}
\usepackage{epsfig}
\usepackage{subfigure}
\usepackage{hhline}
\usepackage{cite}

\newcommand{\reaktion}{\mbox{$\vec{p}p\rightarrow\,pp\:\!\eta$ }}

\begin{document}
\title{Analysing power \unboldmath{$A_y$} in the reaction \reaktion near threshold%
\thanks{Presented at MESON 2002
}%
}
\author{P. Winter for the COSY-11 collaboration
\address{Institut f\"ur Kernphysik, Forschungszentrum J\"ulich, Germany}
}
\maketitle
\begin{abstract}
A first measurement of the analysing power in the reaction $\vec{p}p \rightarrow pp\eta$ near the production threshold has been performed at the internal facility COSY-11 at the COoler SYnchrotron COSY. 
\end{abstract}
  
\section{Introduction}
The $\eta$ meson production in $NN$ collisions has become an active field of studies in nuclear physics within the last years. Several measurements covering more than 100 MeV excess energy have been performed at different accelerators. Total \cite{bergdolt:93, chiavassa:94, calen:96, calen:97, hibou:98, smyrski:00} as well as differential cross sections \cite{calen:99, tatischeff:00, moskal:01-2, abdelbary:02} provide an extensive basis for theoretical investigations to understand the production mechanism on the hadronic and quark-gluon level. Different existing theoretical descriptions describe the data well. Therefore, additional measurements are needed for a qualitative investigation. Polarisation observables, such as the analysing power, present a powerful tool because they are sensitive to the influence of higher partial waves.

\section{Experiment}
\begin{figure}[ht]
\begin{center}
\subfigure[\label{invmass}]{\epsfig{file=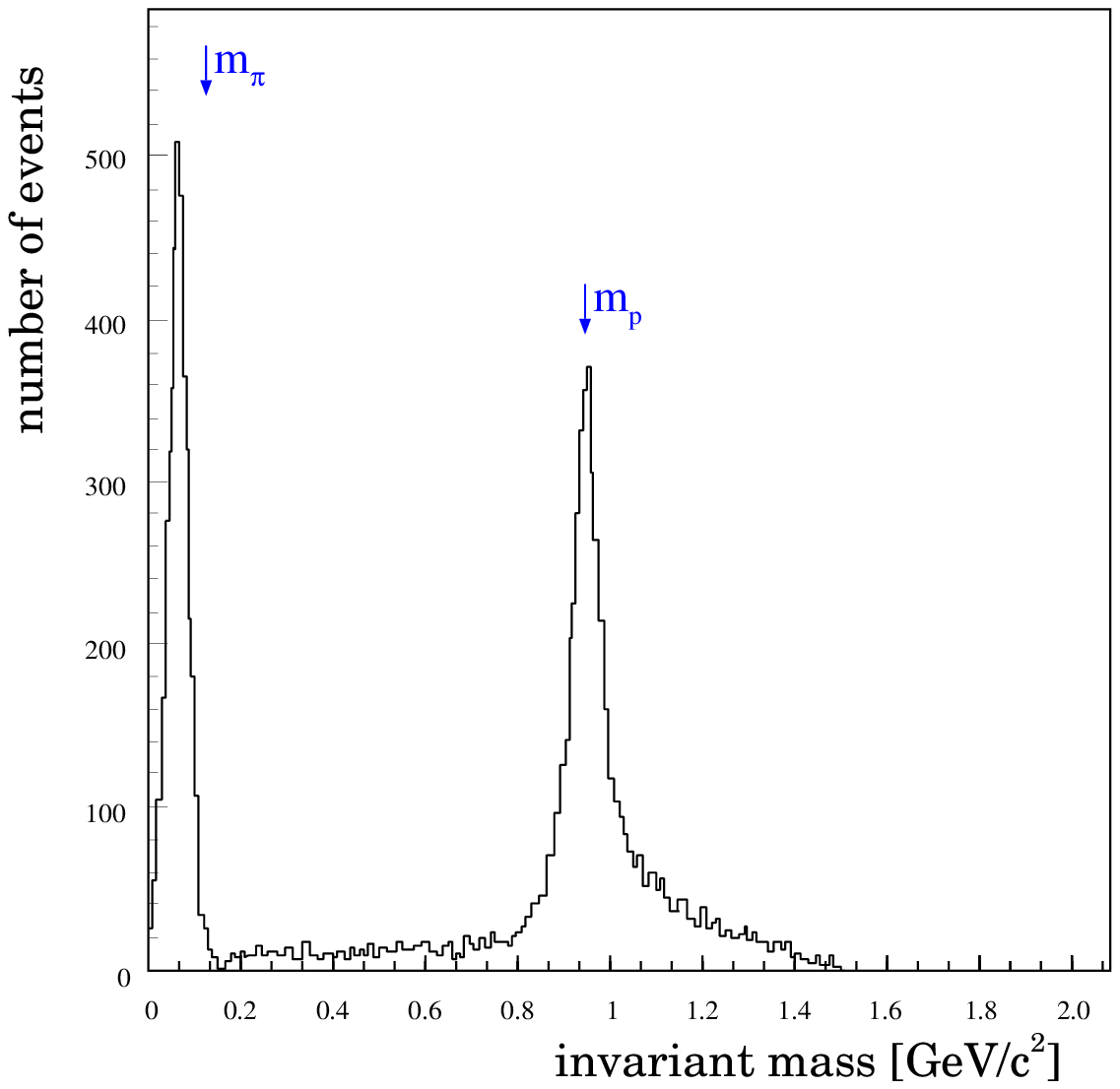,width=0.33\columnwidth}}
\hspace{1cm}
\subfigure[\label{missmass}]{\epsfig{file=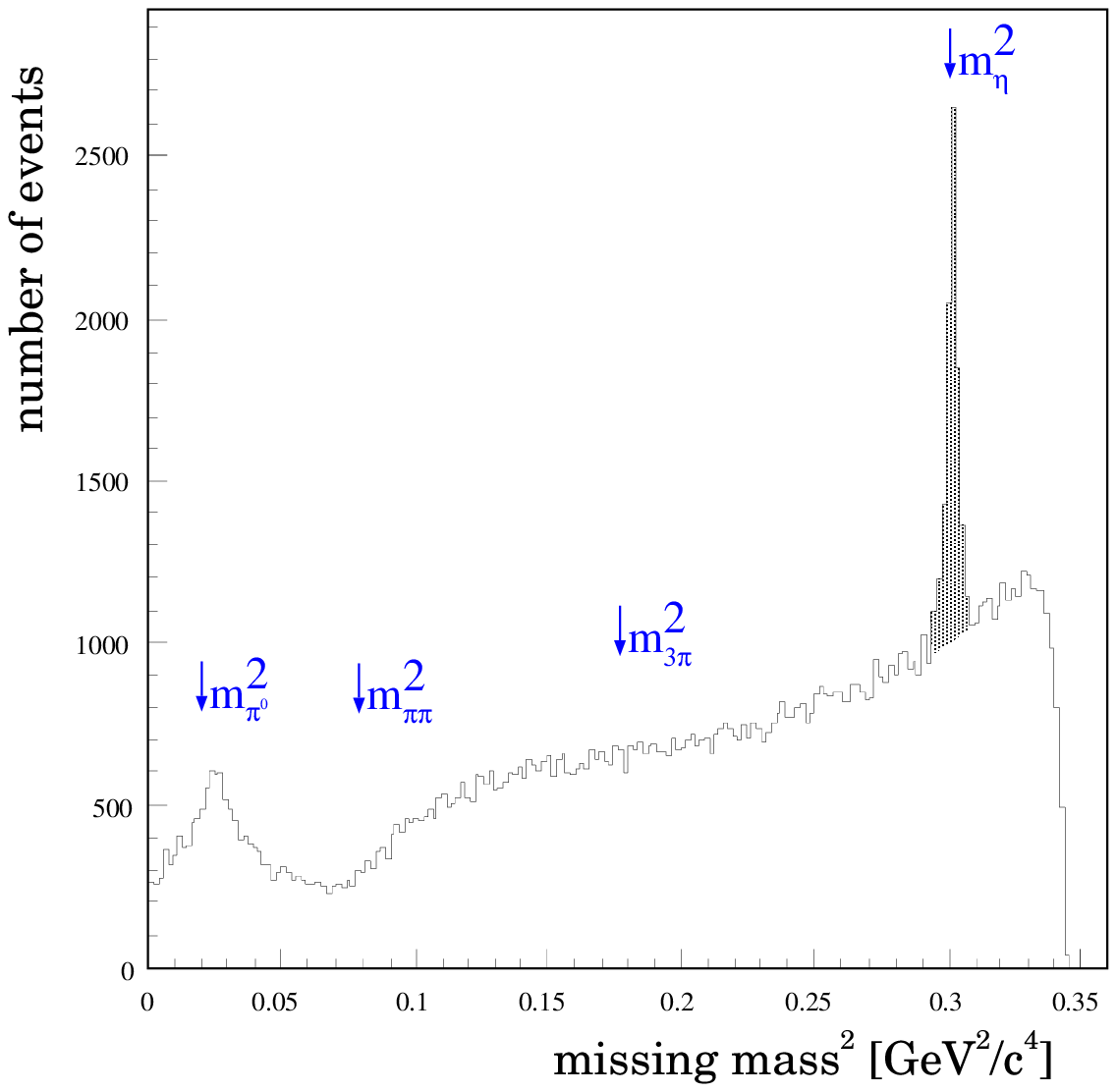,width=0.33\columnwidth}}
\end{center}
\caption{(a) Invariant mass spectrum for events with two reconstructed tracks. Besides the clear proton peak a second signal stemming from pions is observed. (b) Missing mass squared for events with two protons in the exit channel.}
\end{figure}

The measurement of the \reaktion reaction was carried out at the internal experiment COSY-11 \cite{brauksiepe:96} at the COler SYnchrotron COSY \cite{maier:97nim} in J\"ulich. The beam momentum was adjusted to p$_{beam}=2.096\,$GeV/c which equals to an excess energy of $Q=40\,$MeV. Via the reconstruction of the four momentum of positively charged particles with a track reconstruction and a time-of-flight measurement, the invariant mass of these particles is calculated (see Figure \ref{invmass}). A clear separation of events with two protons from events with pions is possible and hence, the not registered $\eta$-meson is identified by means of the missing mass method (Figure \ref{missmass}). Finally, the number of $\eta$-events $N_{\uparrow,\downarrow}$ are extracted. An efficiency correction for the polar and azimuthal angle of the relative proton-proton momentum in the pp-rest system was applied which is necessary for an extraction of partial wave interference terms.

Together with an evident $\eta$ signal there is a physical background mainly resulting from multi pion production. However, subtracting this background enables to determine the number of $\eta$ events $N_{\uparrow,\downarrow}$ for spin up and spin down.

\section{Results}
The determination of the analysing power requires the knowledge of the averaged beam polarisation P$_{\uparrow,\downarrow}$ for the cycles with spin up and down, respectively. Furthermore, the relative time-integrated luminosity $\mathcal{L}_{rel}:=\int \mathcal{L}_\downarrow\,dt_\downarrow\, /\, \int \mathcal{L}_\uparrow\,dt_\uparrow$ and the number of \reaktion-events $N_{\uparrow,\downarrow}$ are required.

The beam polarisation was determined at the internal experiment EDDA \cite{rohdjess:97} with a simultaneous measurement of the elastic pp-scattering to P$_\uparrow=0.439\pm0.007$ and P$_\downarrow=-0.535\pm0.007$. The relative luminosity $\mathcal{L}_{rel}=0.977\pm0.002$ was extracted via the elastic proton-proton scattering. 

\begin{figure}[ht]
\begin{center}
\epsfig{file=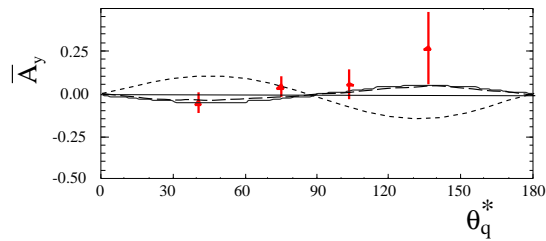,scale=0.77}
\end{center}
\caption{Analysing power as a function of the CMS angle $\theta_q^*$ (triangles). Furthermore calculations from \cite{faeldt:01} (solid line) and \cite{nakayama:02} (dotted and dashed curves) are plotted.\label{ay-vergleich}}
\end{figure}

The analysing power is shown in figure \ref{ay-vergleich} as a function of the center of mass polar angle of the $\eta$-meson with respect to the beam direction together with two predictions based on meson exchange models. The solid line corresponds to calculations of \cite{faeldt:01} based on dominant $\rho$ exchange. The dotted line is a full model calculation of \cite{nakayama:02} while the dashed curve is a reduction to vector meson exchange.
The comparison shows a slight deviation of the theoretical predictions towards backward scattering angles. Qualitatively, the data seem to favour the calculations with dominant vector meson exchange. For a more quantitative statement these first results for the analysing power are not yet precise enough. Therefore, a further beam time with higher statistics is scheduled this year.

\section{Acknowledgements}
This work was partly supported by the European Community -- Access to Research Infrastructure action of the Improving Human Potential Programme.

\bibliography{abbrev,polarisation,general}

\end{document}